\documentclass[pra,twoside,twocolumn,showpacs]{revtex4-1}
\usepackage{amsmath,amsfonts,graphicx}
\usepackage[amssymb]{SIunits}
\usepackage[active]{srcltx}
\usepackage{color}

\newcommand{\ket}[1]{|#1\rangle}
\newcommand{\bra}[1]{\langle#1|}
\newcommand{\bracket}[2]{\langle#1|#2\rangle}

\newcommand{\sg}[1]{\mathsf{#1}}
\renewcommand{\vec}{\textbf}

\SetSymbolFont{symbols}{normal}{OMS}{cmsy}{m}{n}

\begin{document}

\bibliographystyle{apsrev4-1}

\title{Relativistic spin operator and Dirac equation} 

\author{Pawe{\l} Caban}\email{P.Caban@merlin.phys.uni.lodz.pl}
\author{Jakub Rembieli{\'n}ski}\email{jaremb@uni.lodz.pl}
\author{Marta W{\l}odarczyk}

\affiliation{Department of Theoretical Physics, University of Lodz\\
Pomorska 149/153, 90-236 {\L}{\'o}d{\'z}, Poland}

\date{\today}

\begin{abstract}
We give a direct link between description of Dirac particles in the
abstract framework of unitary representation of the 
Poincar\'e group and description with the help of the Dirac equation.
In this context we discuss in detail the spin operator for a relativistic
Dirac particle. 
We show also that the spin operator used in quantum field theory
for spin $s=1/2$ corresponds to the Foldy-Woutheysen mean-spin
operator.
\end{abstract}

 \maketitle

\section{Introduction}

The field of relativistic quantum information theory has emerged
several years ago \cite{Czachor1997_1}. Since then a lot of papers have
been published (see e.g.\ 
\cite{GA2002,AM2002,PST2002,RS2002,CW2003,TU2003_1,LY2004,%
PT2004_1,CR2005,CRW2008,HSZ2007_1,HSZ2008,SHZ2009,Caban2008,%
CDKO_2010_fermion_helicity,PT2003_2,CR2003_Wigner,Caban2007_photons,%
CR2006,LM2009,CDKO_2010_fermion_helicity,FBHH2010,%
PVD2011,CHK2011,SV2011,SV2012,DV2012,CRW2009,Czachor2010,CRWW_2011_hybrid,%
CRW_2011_tensor,CR_2011_rel_correl})
and different aspects of relativistic quantum 
information have been studied (including, for example, correlations in
vector boson systems \cite{CRW2008}, correlations of massive particles
in helicity formalism
\cite{HSZ2007_1,HSZ2008,SHZ2009,Caban2008,CDKO_2010_fermion_helicity}
or massless particles
\cite{TU2003_1,PT2003_2,CR2003_Wigner,Caban2007_photons}). 

However, the spin-1/2 massive particles are often considered as the
best objects to study relativistic effects on entanglement and
violation of Bell-type inequalities.
They have been considered in many papers 
\cite{Czachor1997_1,PST2002,CW2003,CR2005,CR2006,LM2009,%
CDKO_2010_fermion_helicity,FBHH2010}. The most recent ones
\cite{PVD2011,CHK2011,SV2011,SV2012,DV2012} also discuss
relativistic effects in a system of spin-1/2 massive particles.

In some papers such particles are described in the framework of
unitary representations of the Poincar\'e group
(e.g.~\cite{Czachor1997_1,SV2012}) while other 
authors use Dirac equation (e.g.~\cite{CHK2011,DV2012}).

In the present paper we give a direct link between these two
approaches. To this end we formulate Dirac formalism in an abstract
Hilbert space which is a carrier space of an unitary representation of
the Poincar\'e group. To include negative energy solutions of the
Dirac equation we consider as a Hilbert space the direct sum of
carrier spaces of positive and negative energy unitary representations
of the Poincar\'e group for a massive spin-1/2 particle. In this Hilbert
space there exists the standard basis (we call it spin basis) defined
in the context of unitary representations of the Poincar\'e group.
Next we introduce basis which under Lorentz transformations transforms
in a manifestly covariant manner according to the bispinor
representation of the Lorentz group. Vectors of the covariant basis in
a natural way fulfill the Dirac equation.
Thus, in our
approach the Dirac equation is a consequence of the demand of
manifest covariance and form of the bispinor representation.
We show also that the well-known Foldy-Woutheysen transformation, which
diagonalizes the Dirac Hamiltonian, corresponds to the transformation
between covariant and spin basis.

Another important issue in the context of relativistic quantum
information theory is the problem of defining a proper spin observable
for a relativistic particle. This problem  has attracted much
attention in the recent years 
\cite{Czachor1997_1,Terno2003,CR2005,CR2006,CRW2009,Czachor2010}. 
Different propositions of a relativistic spin are still discussed in
the literature \cite{SV2012,DV2012}. Unfortunately, there are also
papers which are 
not free from misunderstandings \cite{CHK2011}.
In particular, in this paper we show that the spin operator we used in
our previous 
papers (e.g. \cite{CR2005,CR2006}) is in fact, for spin $s=1/2$, equal
to the so called mean-spin 
operator defined by Foldy and Woutheysen \cite{FW1950} for a Dirac
particle. We also discuss the transformation properties of this spin
operator under Lorentz group action. We show that the spin operator in
the momentum representation 
transforms according to a Wigner rotation.

We hope that our formalism will be helpful in clarifying some issues
in the field of relativistic quantum information theory.


\section{Relativistic description of a Dirac particle}

\subsection{Dirac equation}

Relativistic spin-1/2 particle is described by
Dirac equation which has the following form
 \begin{equation}
 (i\gamma^\mu \partial_\mu - m)\psi(x) = 0,
 \label{Dirac_eq}
 \end{equation}
where bispinor $\psi(x)$ is a four-component column and we have used
the standard notation 
$\partial_\mu=\frac{\partial}{\partial x^\mu}$. 
For the conventions concerning Dirac matrices and metric tensor see
Appendix~\ref{app:Dirac_matrices}. We use natural units with
$\hbar=c=1$. 

Dirac equation (\ref{Dirac_eq}) is covariant under Lorentz
transformations $\Lambda$
 \begin{equation} 
 {x^\prime}^\mu = \Lambda^{\mu}_{\phantom{\mu}\nu} x^\nu.
 \label{Lorentz_transf}
 \end{equation}
On the level of Dirac bispinors Lorentz transformations are realized as
 \begin{equation}
 \psi^\prime(x^\prime) = S(\Lambda) \psi(x),
 \label{S_Lambda_def}
 \end{equation}
where $S(\Lambda)$ is a 4-dimensional, bispinor representation of the
Lorentz group and $x^\prime=\Lambda x$. 
Bispinor representation is generated by $\Sigma^{\mu\nu}=
\frac{i}{4}[\gamma^\mu,\gamma^\nu]$:
 \begin{equation}
 S(\Lambda) = \exp(i\tfrac{\omega_{\mu\nu}}{2}\Sigma^{\mu\nu}), 
 \label{bispinor_rep_def}
 \end{equation}
therefore
 \begin{equation}
 S^{-1}(\Lambda) = \gamma^0 S^\dag(\Lambda)
 \gamma^0. 
 \label{bispinor_rep_inverse}
 \end{equation}

Covariance of the Dirac
equation gives the following condition: 
 \begin{equation}
 S^{-1}(\Lambda) \gamma^\mu S(\Lambda) =
 \Lambda^{\mu}_{\phantom{\mu}\nu} \gamma^\nu.
 \label{covariance_gamma}
 \end{equation}

We define the invariant scalar product of two Dirac spinors $\psi(x)$
and $\phi(x)$ at time $x^0$ by
 \begin{equation}
 (\psi,\phi) = \int_{x^0=const} d^3\vec{x}\, \overline{\psi}(x) \gamma^0
 \phi(x),
 \label{scalar_spinors_x0}
 \end{equation}
where $\overline{\psi}(x)=\psi^\dag(x)\gamma^0.$
This scalar product can be written in a manifestly invariant form
 \begin{equation}
 (\psi,\phi) = \int d \sigma_\mu(x)\, \overline{\psi}(x) \gamma^\mu
 \phi(x),
 \label{scalar_spinors_general}
 \end{equation}
where integration is performed over a space-like surface $\sigma$ and
$d\sigma_\mu(x) = -
\frac{1}{3!}\varepsilon_{\mu\nu\sigma\lambda} dx^\nu \wedge dx^\sigma
\wedge dx^\lambda$.

\subsection{Unitary representations of the 
Poincar\'e group for a Dirac particle}

As is well known, the Dirac equation possesses
negative energy solutions as well as positive energy ones.
Therefore, to describe in a consistent manner both types of solutions 
as a space of states for a spin-1/2 relativistic particle
we take the direct sum of carrier spaces of two unitary, irreducible
representations of the Poincar\'e group,
$\mathcal{H}=\mathcal{H}_+\oplus\mathcal{H}_-$, 
corresponding to positive and negative energy, respectively.

The carrier space of a unitary irreducible representation
corresponding to the massive, spin-1/2 particle,
$\mathcal{H}_{\sg{\epsilon}}$
($\sg{\epsilon}=+1$ corresponds to
positive energy while $\sg{\epsilon}=-1$ corresponds to negative energy), is
spanned by the 
eigen-vectors of the four-momentum operators $\hat{P}^\mu$
 \begin{equation}
 \hat{P}^\mu \ket{\sg{\epsilon}p,\sigma} =
  \sg{\epsilon}p^\mu \ket{\sg{\epsilon}p,\sigma},
 \label{four-momentum_eigenvectors}
 \end{equation}
where $\sigma=\pm1/2$. The above choice of basis is explained in
Appendix~\ref{App:basis}. 

We denote by $U(\Lambda)$ an unitary operator representing a Lorentz
transformation $\Lambda$. It holds
 \begin{equation}
 U(\Lambda) = \exp\big( i \tfrac{\omega_{\mu\nu}}{2} \hat{J}^{\mu\nu}
 \big), 
 \label{unitary_rep_generators}
 \end{equation}
where $\hat{J}^{\mu\nu}$ are generators of the Lorentz group.

Action of $U(\Lambda)$ on the basis vectors
reads [for derivation see Appendix \ref{App:basis}]:
 \begin{equation}
 U(\Lambda) \ket{\sg{\epsilon}p,\sigma} =
 \mathcal{D}(R(\Lambda,p))_{\lambda\sigma}  
 \ket{\sg{\epsilon}\Lambda p,\lambda},
 \label{transformation}
 \end{equation}
where $\mathcal{D}$ is the standard unitary spin-1/2 representation of the
rotation group, Wigner rotation $R(\Lambda,p)=L^{-1}_{\Lambda
  p}\Lambda L_p$ and $L_p$ denotes standard Lorentz transformation
which is defined by the conditions: $L_pq=p$, $L_q=I$ with
$q=(m,\vec{0})$. 

Let us consider the parity operation $\mathbb{P}$:
$\mathbb{P}(p^0,\vec{p}) = (p^0,-\vec{p})\equiv p^\pi$. Action of
parity on basis vectors reads
 \begin{equation}
 U(\mathbb{P}) \ket{\sg{\epsilon}p,\sigma} =
 \mathsf{P}^{\sg{\epsilon}}_{\sigma\lambda}  
 \ket{\sg{\epsilon}p^{\pi},\lambda}.
 \label{parity_general_basis}
 \end{equation}
Consistency of Eqs.~(\ref{transformation}) and (\ref{parity_general_basis})
leads to
 \begin{equation}
 \mathsf{P}^{\sg{\epsilon}}_{\sigma\lambda} = \xi^{\sg{\epsilon}}
 \delta_{\sigma\lambda},
 \end{equation}
where $|\xi^{\sg{\epsilon}}|=1$. Therefore, finally
 \begin{equation}
 U(\mathbb{P}) \ket{\sg{\epsilon}p,\sigma} = \xi^{\sg{\epsilon}}
 \ket{\sg{\epsilon}p^{\pi},\sigma}.
 \label{parity_final_basis}
 \end{equation}

We adopt the following, Lorentz-covariant normalization of the basis
vectors: 
 \begin{equation}
 \bracket{\sg{\epsilon}^\prime p^\prime,\sigma^\prime}{\sg{\epsilon}p,\sigma} = 2
 \omega(\vec{p}) \delta^3(\vec{p}^\prime-\vec{p})
 \delta_{\sg{\epsilon}\sg{\epsilon}^\prime}
 \delta_{\sigma\sigma^\prime},
 \label{basis_cov}
 \end{equation}
where $\omega(\vec{p})=\sqrt{\vec{p}^2+m^2}$. 

We can define in a natural way a hermitian operator $\hat{\sg{\mathcal{E}}}$
with eigen-values $\sg{\epsilon}=\pm1$ and corresponding eigen-vectors
$\ket{\sg{\epsilon}p,\sigma}$:
 \begin{equation}
 \hat{\sg{\mathcal{E}}}=\frac{\hat{P}^0}{\omega(\vec{p})}.
 \label{C_def}
 \end{equation}
Indeed, Eq.~(\ref{four-momentum_eigenvectors}) imply
 \begin{equation}
 \hat{\sg{\mathcal{E}}} \ket{\sg{\epsilon}p,\sigma} =
 \sg{\epsilon} \ket{\sg{\epsilon}p,\sigma}.
 \end{equation}
Let us notice that operator corresponding to
absolute value of energy can be written as
 \begin{equation}
 |\hat{P}^0| = \sqrt{m^2+\hat{\vec{P}}^2} = \hat{\sg{\mathcal{E}}} \hat{P}^0.
 \label{energy_abs_value_C}
 \end{equation}
We will use this observation later on.

One can also check that the following resolution of unity holds
 \begin{equation}
 \openone = \sum_{\sg{\epsilon}}\sum_{\sigma}
 \int \frac{d^3\vec{p}}{2\omega(\vec{p})}
 \ket{\sg{\epsilon}p,\sigma}\bra{\sg{\epsilon}p,\sigma},
 \label{jedynka_p_sigma}
 \end{equation}
where the measure $d^3\vec{p}/(2\omega(\vec{p}))$ is Lorentz-invariant.

\subsection{Manifestly covariant basis}

The spin basis $\{\ket{\sg{\epsilon}p,\sigma}\}$, although naturally
defined in the framework of unitary representation of Poincar\'e group,
is not manifestly covariant [Eq.~(\ref{transformation})].

Therefore, we define another basis in the space $\mathcal{H}$
 \begin{equation}
 \ket{\alpha,\sg{\epsilon}p} = \sum_{\sigma} v_{\alpha\sigma}^{\sg{\epsilon}}(p) 
 \ket{\sg{\epsilon}p,\sigma},
 \label{covariant_basis_def}
 \end{equation}
where we demand
 \begin{equation}
 U(\Lambda) \ket{\alpha,\sg{\epsilon}p} =
 S^{-1}(\Lambda)_{\alpha\beta} 
 \ket{\beta,\sg{\epsilon}\Lambda p}. 
 \label{transformation_covariant}
 \end{equation}
In Eq.~(\ref{transformation_covariant}) $S(\Lambda)$ denotes
the bispinor representation of the Lorentz group [compare
Eq.~(\ref{S_Lambda_def})]\footnote{Notice that in our previous papers
  \cite{CR2005,CR2006} we have 
denoted the bispinor representation as $\mathsf{D}(\Lambda)$. However,
in textbooks dealing with Dirac equation, like \cite{BD1964}, the letter $S$
is used in this context.}. Therefore, $\alpha$ is a bispinor index.

Inserting $\Lambda=\mathbb{P}$ in Eq.~(\ref{covariance_gamma}), we can
uniquely determine bispinor representation of the parity operator
 \begin{equation}
 S(\mathbb{P}) = \xi \gamma^0,
 \label{parity_def}
 \end{equation}
where $|\xi|=1$. Therefore
 \begin{equation}
 U(\mathbb{P}) \ket{\alpha,\sg{\epsilon}p} = \xi^* \sum_\beta 
 \gamma^{0}_{\alpha\beta} \ket{\beta,\sg{\epsilon}p^\pi}.
 \label{parity_on_covariant}
 \end{equation}
Of course, phase factors $\xi$ [Eq.~(\ref{parity_def})] and
$\xi^{\sg{\epsilon}}$ [Eq.~(\ref{parity_final_basis})] are related.
Indeed, by virtue of Eqs.~(\ref{parity_final_basis},
\ref{covariant_basis_def}, \ref{parity_on_covariant}) we obtain
 \begin{equation}
 \xi^{\sg{\epsilon}} v^{\sg{\epsilon}}(p) = \xi^* \gamma^0 v^{\sg{\epsilon}}(p^\pi),
 \label{xi_xi_c_1}
 \end{equation}
where $v^{\sg{\epsilon}}(p)=[v^{\sg{\epsilon}}_{\alpha\sigma}(p)]$.

Eqs.~(\ref{transformation}, \ref{transformation_covariant}) imply the
following Weinberg consistency condition:
 \begin{equation}
 S(\Lambda) v^{\sg{\epsilon}}(p)\mathcal{D}^T(R(\Lambda,p)) = 
 v^{\sg{\epsilon}}(\Lambda p).
 \label{Weinberg_cond}
 \end{equation}
By virtue of the above equation we have
 \begin{equation}
 v^{\sg{\epsilon}}(p)=S(L_p) v^{\sg{\epsilon}}(q),
 \label{amplitudes_generation}
 \end{equation}
where $q=(m,\vec{0})$ is the four-momentum in the rest frame of a
particle. 
The Weinberg condition (\ref{Weinberg_cond}) and
Eq.~(\ref{bispinor_rep_inverse}) imply
 \begin{align}
 \mathcal{D}^*(R(\Lambda,p)) \bar{v}^{\sg{\epsilon}^\prime}(p) & v^{\sg{\epsilon}}(p)
 \mathcal{D}^T(R(\Lambda,p))  = 
 \bar{v}^{\sg{\epsilon}^\prime}(\Lambda p) v^{\sg{\epsilon}}(\Lambda p),
 \label{Weinberg_bar_v_v}\\
 S(\Lambda) v^{\sg{\epsilon}}(p) & \bar{v}^{\sg{\epsilon}^\prime}(p)
 S^{-1}(\Lambda)  = v^{\sg{\epsilon}}(\Lambda p) 
 \bar{v}^{\sg{\epsilon}^\prime}(\Lambda p),
 \label{Weinberg_v_bar_v}
 \end{align}
where $\bar{v}^{\sg{\epsilon}}(p) \equiv {v^{\sg{\epsilon}}}^\dag(p) \gamma^0$.

Now, Eqs.~(\ref{parity_def}, \ref{amplitudes_generation},
\ref{Weinberg_bar_v_v}, \ref{Weinberg_v_bar_v},
\ref{bispinor_rep_inverse}) and Schur's Lemma lead to  
 \begin{equation}
 v^{\sg{\epsilon}}(p) \bar{v}^{\sg{\epsilon}}(p) = \sg{\epsilon} \Lambda_{\sg{\epsilon}}(p),
 \label{v_bar_v_2}
 \end{equation}
where we have introduced the standard projectors
 \begin{equation}
 \Lambda_{\sg{\epsilon}}(p)=
 \frac{mI+\sg{\epsilon}p\gamma}{2m},
 \label{projectors_Lambda}
 \end{equation}
and
 \begin{equation}
 \bar{v}^{\sg{\epsilon}^\prime}(p) v^{\sg{\epsilon}}(p) = \sg{\epsilon}
 \delta_{\sg{\epsilon}\sg{\epsilon}^\prime} I_2.
 \label{bar_v_v_2}
 \end{equation}
Eq.~(\ref{v_bar_v_2}) imply the following condition:
 \begin{equation}
 \sum_{\sg{\epsilon}} \sg{\epsilon} v^{\sg{\epsilon}}(p) \bar{v}^{\sg{\epsilon}}(p) = I.
 \end{equation}
The explicit form of amplitudes $v^{\sg{\epsilon}}(p)$ can be easily
determined with help of Eqs.~(\ref{amplitudes_generation},
\ref{v_bar_v_2}, \ref{bar_v_v_2}, \ref{gamma_explicit}) and is given
in Appendix~\ref{sec:formulas} [Eq.~(\ref{amplitudes_explicit})].

Now, using Eqs.~(\ref{bar_v_v_2}, \ref{amplitudes_explicit}) we can
simplify Eq.~(\ref{xi_xi_c_1}). We get finally
 \begin{equation}
 \xi^\sg{\epsilon} = \sg{\epsilon} \xi^*.
 \end{equation}

The covariant basis vectors fulfill the following normalization
condition:
 \begin{equation}
 \bracket{\overline{\beta,\sg{\epsilon}^\prime k}}{\alpha,\sg{\epsilon}p} = 
 2 \sg{\epsilon} \omega(\vec{p}) \delta_{\sg{\epsilon}^\prime\sg{\epsilon}} 
 \delta^3(\vec{k}-\vec{p}) 
 \big(\Lambda_{\sg{\epsilon}}(p) \big)_{\alpha\beta},
 \label{cov_basis_normalization}
 \end{equation}
where we have used the natural notation
 \begin{equation}
 \bra{\overline{\beta,\sg{\epsilon}k}} = \sum_\alpha \bra{\alpha,\sg{\epsilon}k}
 \gamma^{0}_{\alpha\beta}.
 \end{equation}
By virtue of Eq.~(\ref{bar_v_v_2}) we 
can invert relation (\ref{covariant_basis_def})
 \begin{equation}
 \ket{\sg{\epsilon}p,\sigma} = \sum_\alpha \sg{\epsilon}
 \bar{v}^{\sg{\epsilon}}_{\sigma\alpha}(p) \ket{\alpha,\sg{\epsilon}p}.
 \label{covariant_basis_inverse}
 \end{equation}
Now, using Eqs.~(\ref{v_bar_v_2}, \ref{covariant_basis_inverse}), we
receive
 \begin{equation}
 \sum_{\beta} (p\gamma)_{\alpha\beta} \ket{\beta,\sg{\epsilon}p} = 
 \sg{\epsilon} m \ket{\alpha,\sg{\epsilon}p},
 \label{pgamma_action}
 \end{equation}
or, in terms of the projectors (\ref{projectors_Lambda})
 \begin{equation}
 \sum_{\beta}\big(\Lambda_{\sg{\epsilon}}(p)\big)_{\alpha\beta}
 \ket{\beta,\sg{\epsilon}p} =  \ket{\alpha,\sg{\epsilon}p}.
 \label{projector_on_cov_basis}
 \end{equation}
Notice that the above equation is in fact the Dirac equation in
momentum representation written in an abstract Hilbert space. Thus, in
this approach the Dirac equation is a consequence of the demand of
manifest covariance and form of the bispinor representation.
Dirac equation (\ref{projector_on_cov_basis}) can be cast in an
operator form
 \begin{equation}
 \sum_\beta \big( \hat{P}\gamma-mI  \big)_{\alpha\beta}
 \ket{\beta,\sg{\epsilon}p} =0.
 \label{Dirac_eq_operator}
 \end{equation}
Therefore, Hamiltonian acts on basis vectors as follows:
 \begin{align}
 \hat{P}^0 \ket{\alpha,\sg{\epsilon}p} & = 
 \sum_\beta \big[ \gamma^0(\sg{\epsilon}\vec{p}\cdot\boldsymbol{\gamma}
 +mI) \big]_{\alpha\beta} \ket{\beta,\sg{\epsilon}p}\nonumber\\
 & \equiv \sum_\beta H^{\sg{\epsilon}}_{D\,\alpha\beta}
 \ket{\beta,\sg{\epsilon}p}. 
 \label{Hamiltonian_action_basis}
 \end{align}

Applying Eq.~(\ref{covariant_basis_inverse}) to
Eq.~(\ref{jedynka_p_sigma}) we get
 \begin{equation}
 \openone = \sum_{\sg{\epsilon}} \sum_{\alpha} 
 \int \frac{d^3\vec{p}}{2\omega(\vec{p})} \sg{\epsilon}
 \ket{\alpha,\sg{\epsilon}p}\bra{\overline{\alpha,\sg{\epsilon}p\phantom{l}}}.
 \end{equation}

\subsection{Functional realization}

In this section we construct a functional realization in terms of
Dirac bispinors. We
consider functional realization with help of covariant
[Eq.~(\ref{covariant_basis_def})] as 
well as spin basis [Eq.~(\ref{four-momentum_eigenvectors})].

\paragraph{Covariant basis}
Let us expand an arbitrary state vector in the covariant basis defined
in Eq.~(\ref{covariant_basis_def})
 \begin{equation}
 \ket{\psi} = \sum_{\sg{\epsilon},\alpha} \int \frac{d^3\vec{p}}{2\omega(\vec{p})} 
 \sg{\epsilon} C_{\psi}(p)^{\sg{\epsilon}}_{\alpha} \ket{\alpha,\sg{\epsilon}p}.
 \label{expansion_cov_basis}
 \end{equation}
Using Eqs.~(\ref{projector_on_cov_basis}) and
(\ref{cov_basis_normalization}) we find 
 \begin{equation}
 \sum_\alpha C_{\psi}(p)^{\sg{\epsilon}}_{\alpha}
 \big(\Lambda_{\sg{\epsilon}}(p)\big)_{\alpha\beta} =
 C_{\psi}(p)^{\sg{\epsilon}}_{\beta}
 \label{coeff_C_Dirac_eq}
 \end{equation}
and
 \begin{equation}
 C_{\psi}(p)^{\sg{\epsilon}}_{\alpha} =
 \bracket{\overline{\alpha,\sg{\epsilon}p\phantom{l}}}{\psi},
 \end{equation}
respectively. 
We would like to connect the function $C_{\psi}(p)^{\sg{\epsilon}}_{\alpha}$
with Dirac bispinor $\psi^{\sg{\epsilon}}_{\alpha}(p)$.
The above equations suggest the following
identification:
 \begin{equation}
 C_{\psi}(p)^{\sg{\epsilon}}_{\alpha} = \overline{\psi}^{\sg{\epsilon}}_{\alpha}(p) = 
 \sum_\beta \psi^{\sg{\epsilon}*}_{\beta}(p) \gamma^{0}_{\beta\alpha}.
 \end{equation}
Therefore, we define a bispinor with definite energy
($\sg{\epsilon}=+1$---positive, $\sg{\epsilon}=-1$---negative) in momentum
representation as follows: 
 \begin{equation}
 \psi^{\sg{\epsilon}}_{\alpha}(p) = \bracket{\psi}{\alpha,\sg{\epsilon}p}.
 \label{bispinor_p_c}
 \end{equation}
Using this definition, Eq.~(\ref{expansion_cov_basis}) takes the
following form 
 \begin{equation}
 \ket{\psi} = \sum_{\sg{\epsilon}} \sum_{\alpha} 
 \int \frac{d^3\vec{p}}{2\omega(\vec{p})}
 \sg{\epsilon} \overline{\psi}^{\sg{\epsilon}}_{\alpha}(p)
 \ket{\alpha,\sg{\epsilon}p}.
 \label{expansion_cov_basis_bispinors}
 \end{equation}
We can easily check that a bispinor (\ref{bispinor_p_c}) fulfills the
Dirac equation (compare Eq.~(\ref{coeff_C_Dirac_eq}))
 \begin{equation}
 \Lambda_{-\sg{\epsilon}}(p) \psi^{\sg{\epsilon}}(p)=0,
 \label{Dirac_eq_bispinor}
 \end{equation}
where $\psi^{\sg{\epsilon}}(p)$ denotes four-component column
$[\psi^{\sg{\epsilon}}_{\alpha}(p)]$. 
 
We can also check that bispinors defined in Eq.~(\ref{bispinor_p_c})
transform properly under Lorentz transformations.
Denoting
 \begin{equation}
 \psi^{\prime\sg{\epsilon}}_{\alpha}(p^\prime) =
 \bracket{\psi^\prime}{\alpha,\sg{\epsilon}p^\prime},
 \end{equation}
where
 \begin{equation}
 U(\Lambda) \ket{\psi} = \ket{\psi^\prime}, \qquad p^\prime=\Lambda p,
 \end{equation}
we receive the following transformation law: 
 \begin{equation}
 \psi^{\prime\sg{\epsilon}}(p^\prime) =
 S(\Lambda) \psi^{\sg{\epsilon}}(p),
 \end{equation}
[compare Eq.~(\ref{S_Lambda_def})]. 
So, we can define the  most general bispinor in momentum
representation as follows 
 \begin{equation}
 \psi_\alpha(p) = \sum_{\sg{\epsilon}} \psi^{\sg{\epsilon}}_{\alpha}(p).
 \end{equation}

Finally, action of the Hamiltonian operator on a bispinor
$\psi^{\sg{\epsilon}}(p)$ can be determined with help of
Eqs.~(\ref{Hamiltonian_action_basis},
\ref{expansion_cov_basis_bispinors}, \ref{Dirac_eq_bispinor}). We get
 \begin{equation}
 \hat{P}^0 \psi^{\sg{\epsilon}}_{\alpha}(p) =
 \sum_\beta \big[ \gamma^0(\sg{\epsilon}\vec{p}\cdot\boldsymbol{\gamma}
 +mI) \big]_{\alpha\beta}
 \psi^{\sg{\epsilon}}_{\beta}(p).
 \label{Hamiltonian_bispinor_covariat}
 \end{equation}

We define scalar product of bispinors in the
following way:
 \begin{equation}
 (\psi,\phi) = \bracket{\phi}{\psi} = \sum_{\sg{\epsilon}}
 \int \frac{d^3\vec{p}}{2\omega(\vec{p})} 
 \sg{\epsilon} \overline{\psi}^{\sg{\epsilon}}(p) \phi^{\sg{\epsilon}}(p).
 \label{scalar_bispinor_p}
 \end{equation}

\paragraph{Spin basis}
Of course, the expansion given in Eq.~(\ref{expansion_cov_basis}) can
be performed in the non-covariant (spin) basis
$\{\ket{\sg{\epsilon}p,\sigma}\}$, too. If, in analogy to
Eq.~(\ref{bispinor_p_c}), we denote 
 \begin{equation}
 \widetilde{\psi}^{\sg{\epsilon}}_{\sigma}(p) = 
 \bracket{\psi}{\sg{\epsilon}p,\sigma},
 \label{spinor_p_c_sigma}
 \end{equation}
then
 \begin{equation}
 \ket{\psi} = \sum_{\sg{\epsilon},\sigma} \int 
 \frac{d^3\vec{p}}{2\omega(\vec{p})} 
 \widetilde{\psi}^{\sg{\epsilon}*}_{\sigma}(p) 
 \ket{\sg{\epsilon}p,\sigma}.
 \label{expansion_spin_basis}
 \end{equation}

In terms of spinors defined in Eq.~(\ref{spinor_p_c_sigma}), the 
scalar product defined in Eq.~(\ref{scalar_bispinor_p}) takes the
form: 
 \begin{equation}
 (\psi,\phi) = \sum_{\sg{\epsilon}}
 \int \frac{d^3\vec{p}}{2\omega(\vec{p})} 
 \widetilde{\psi}^{\sg{\epsilon}\dag}(p) \widetilde{\phi}^{\sg{\epsilon}}(p).
 \end{equation}

\paragraph{Relation between bases}
Spinors $\psi^{\sg{\epsilon}}_{\alpha}(p)$ and
$\widetilde{\psi}^{\sg{\epsilon}}_{\sigma}(p)$ are related via the following
relation:
 \begin{equation}
 \psi^{\sg{\epsilon}}_{\alpha}(p) = \sum_{\sigma}
 v^{\sg{\epsilon}}_{\alpha\sigma}(p)
 \widetilde{\psi}^{\sg{\epsilon}}_{\sigma}(p),
 \label{bases_relation}
 \end{equation}
where we have used Eq.~(\ref{covariant_basis_def}).

\section{Newton-Wigner position operator}
\label{sec:position}

Problem of defining a proper position operator in the relativistic
quantum mechanics has a very long history and no fully satisfactory
solution (1see e.g.,~Ref.~\cite{cab_Bacry1988}).
In this section we discuss briefly the Newton--Wigner position
operator \cite{cab_NW1949} which, although non-covariant, seems to be
the best proposition.  
The Newton--Wigner position operator is assumed to be hermitian, to
have commuting components 
 \begin{equation}
 [\hat{X}^i,\hat{X}^j] = 0,
 \label{NWXiXj}
 \end{equation}
and to fulfill standard canonical commutation relations with
four-momentum operators
 \begin{equation}
 [\hat{X}^i,\hat{P}^j] = i\delta^{ij}.
 \label{NW_momentum_commutator}
 \end{equation}
Eqs.~(\ref{four-momentum_eigenvectors}, \ref{NW_momentum_commutator})
imply the following relation
 \begin{equation}
 e^{i\vec{a}\cdot\hat{\vec{X}}} \ket{\sg{\epsilon}p,\sigma} = N(p,\sg{\epsilon}\vec{a}) 
 \ket{\sg{\epsilon}p(\sg{\epsilon}\vec{a}),\sigma},
 \label{exp_X_action_spin_basis}
 \end{equation}
where we have denoted by $p(\sg{\epsilon}\vec{a})$ a four-vector with components
given below
 \begin{equation}
 p^0(\sg{\epsilon}\vec{a})= \omega(\vec{p}+\sg{\epsilon}\vec{a})=
 \sqrt{m^2+(\vec{p}+\sg{\epsilon}\vec{a})^2},\quad
 \vec{p}(\sg{\epsilon}\vec{a}) = \vec{p}+\sg{\epsilon}\vec{a},
 \end{equation}
and the normalization factor $N(p,\sg{\epsilon}\vec{a})$ is equal to
 \begin{equation}
 N(p,\sg{\epsilon}\vec{a}) = 
 \sqrt{\frac{\omega(\vec{p})}{\omega(\vec{p}+\sg{\epsilon}\vec{a})}} 
 = \left( \frac{\vec{p}^2+m^2}{(\vec{p}+\sg{\epsilon}\vec{a})^2+m^2}
 \right)^{1/4}. 
 \label{Npca}
 \end{equation}
Eqs.~(\ref{exp_X_action_spin_basis}, \ref{Npca}) imply the well-known
relation  
 \begin{equation}
 \hat{\vec{X}} \widetilde{\psi}^{\sg{\epsilon}}_{\sigma}(p) =  
 i \sg{\epsilon} \left( \nabla_{\vec{p}} - \frac{1}{2} 
 \frac{\vec{p}}{\vec{p}^2+m^2} \right) 
 \widetilde{\psi}^{\sg{\epsilon}}_{\sigma}(p).
 \end{equation}

Using Eqs.~(\ref{covariant_basis_def}, \ref{covariant_basis_inverse},
\ref{exp_X_action_spin_basis}) 
we find in a bispinor (covariant) basis
 \begin{multline}
 e^{i\vec{a}\cdot\hat{\vec{X}}} \ket{\alpha,\sg{\epsilon}p} = 
 \sg{\epsilon} N(p,\sg{\epsilon}\vec{a})\\
 \times \sum_{\beta} 
 \big( v^{\sg{\epsilon}}(p) \bar{v}^{\sg{\epsilon}}(p(\sg{\epsilon}\vec{a}))
 \big)_{\alpha\beta}  
 \ket{\beta,\sg{\epsilon}p(\sg{\epsilon}\vec{a})}.
 \end{multline}
Therefore,
for wave functions in a bispinor basis, defined in
Eq.~(\ref{bispinor_p_c}), we get
 \begin{multline}
 \hat{\vec{X}} \psi^{\sg{\epsilon}}_{\alpha}(p) = 
 -i \sum_\beta \Big(\big( \nabla_{\vec{p}} v^{\sg{\epsilon}}(p)
 \big) \bar{v}^{\sg{\epsilon}}(p)\Big)_{\alpha\beta}
 \psi^{\sg{\epsilon}}_{\beta}(p) \\
  + i \sg{\epsilon} \Big( \nabla_{\vec{p}} - 
 \frac{1}{2} \frac{\vec{p}}{\vec{p}^2+m^2}\Big)
 \psi^{\sg{\epsilon}}_{\alpha}(p).
 \end{multline}

\section{The Foldy-Woutheysen transformation}
\label{sec:FW}

The Foldy-Woutheysen (FW) transformation \cite{FW1950}
is a canonical transformation  which diagonalizes Dirac Hamiltonian
given in Eq.~(\ref{Hamiltonian_action_basis}) or
(\ref{Hamiltonian_bispinor_covariat}).
Hamiltonian (\ref{Hamiltonian_action_basis}) is defined in the
covariant basis. Using Eqs.~(\ref{bar_v_v_2}, \ref{bases_relation}) we
can find Hamiltonian in the spin basis. We have
 \begin{equation}
 \hat{P}^0 \widetilde{\psi}^{\sg{\epsilon}}_{\lambda} =
 \sum_{\alpha\beta} \sg{\epsilon}
 \bar{v}^{\sg{\epsilon}}_{\lambda\alpha}(p)
 H^{\sg{\epsilon}}_{D\,\alpha\beta}  v^{\sg{\epsilon}}_{\beta\sigma} 
 \widetilde{\psi}^{\sg{\epsilon}}_{\sigma}.
 \end{equation}
Therefore, by virtue of Eqs.~(\ref{formula_1}, \ref{formula_5}) we
finally get 
 \begin{equation}
 \hat{P}^0 \widetilde{\psi}^{\sg{\epsilon}}_{\lambda} = 
 \sg{\epsilon} p^0 \widetilde{\psi}^{\sg{\epsilon}}_{\lambda}. 
 \end{equation}
Thus we see that Foldy-Woutheysen transformation corresponds to change
of basis from manifestly covariant one to spin one. Foldy-Woutheysen
spinors are simply spinors defined in terms of vectors spanning the
carrier space of the unitary representation of the Poincar\'e group.

\section{Spin operator}

In this section we clarify some questions concerning spin operator for
a Dirac particle.   

Spin is an internal degree of freedom. It means that a spin operator
should commute with space-time observables like momentum and
position. 
Therefore, choosing as a position operator the Newton--Wigner one,
which fulfills the relations (\ref{NWXiXj},
\ref{NW_momentum_commutator}), we find that the
action of a spin component operator, $\hat{S}^i$,
on a spin basis vectors $\{\ket{\sg{\epsilon}p,\sigma}\}$ must have the
following form:
 \begin{equation}
 \hat{S}^i \ket{\sg{\epsilon}p,\sigma} = \sum_\lambda A_{\sigma\lambda}^{i} 
 \ket{\sg{\epsilon}p,\lambda},
 \end{equation}
where $[A_{\sigma\lambda}^{i}]$ are constant $2\times2$ matrices.
Moreover, we demand that spin components fulfill standard commutation
relation
 \begin{equation}
 [\hat{S}^i,\hat{S}^j] = i \varepsilon_{ijk} \hat{S}^k.
 \label{spin_components_commutation}
 \end{equation}
Thus, we are lead to
 \begin{equation}
 \hat{S}^i \ket{\sg{\epsilon}p,\sigma} = \frac{1}{2} \sum_\lambda
 (\sigma_{i}^{T})_{\sigma\lambda}  
 \ket{\sg{\epsilon}p,\lambda},
 \label{spin_action_spin_basis}
 \end{equation}
where $\sigma_i$ are standard Pauli matrices.

On the other hand, we can try to define spin operator in terms of the
generators of the Poincar\'e group. We know that spin square operator
is well defined in the unitary representation of the Poincar\'e group
and has the following form
 \begin{equation}
 \hat{\vec{S}}^2 = - \frac{1}{m^2} \hat{W}^\mu \hat{W}_\mu,
 \end{equation}
where $\hat{W}^\mu$ is the Pauli-Lubanski (pseudo)four-vector
 \begin{equation}
 \hat{W}^\mu =  \frac{1}{2} \varepsilon^{\nu\alpha\beta\mu} \hat{P}_\nu
 \hat{J}_{\alpha\beta},
 \label{Pauli-Lubanski_def}
 \end{equation}
and $\hat{J}_{\alpha\beta}$ are generators of the Lorentz
group. Therefore, taking into account that spin is a pseudo-vector, it
is natural to look for a spin operator which is a 
linear function of components of $\hat{W}^\mu$. 
If we assume that a spin operator
is such a function and
(i) commutes with four-momentum operators, (ii) fulfills the canonical
commutation relations (\ref{spin_components_commutation}), (iii)
transforms like a vector under rotations, i.e.
 \begin{equation}
 [\hat{J}^i,\hat{S}^j] = i \varepsilon_{ijk} \hat{S}^k,
 \label{spin_transf_rotat_algebra}
 \end{equation}
where $\hat{J}^i = \frac{1}{2}\varepsilon_{ijk}\hat{J}^{jk}$, we
arrive at
 \begin{equation}
 \hat{\vec{S}} = \frac{1}{m} \left( \frac{|\hat{P}^0|}{\hat{P}^0}
 \hat{\vec{W}} - \hat{W}^0
   \frac{\hat{\vec{P}}}{|\hat{P}^0|+m} \right).
 \label{spin_NW_def_abs}
 \end{equation}
With help of Eq.~(\ref{energy_abs_value_C}) we finally get
 \begin{equation}
 \hat{\vec{S}} = \frac{1}{m} \left(
   \hat{\sg{\mathcal{E}}}\hat{\vec{W}} - \hat{W}^0 
   \frac{\hat{\vec{P}}}{\hat{\sg{\mathcal{E}}}\hat{P}^0+m} \right).
 \label{spin_NW_def_C}
 \end{equation}
Let us stress that in the case when only positive energies are
allowed, the spin operator given in Eq.~(\ref{spin_NW_def_C}) takes
the well-known form
 \begin{equation}
 \hat{\vec{S}} = \frac{1}{m} \left( \hat{\vec{W}} - \hat{W}^0
   \frac{\hat{\vec{P}}}{\hat{P}^0+m} \right).
 \label{spin_NW_def_positive}
 \end{equation}
This spin operator is used in quantum field theory (see e.g.\
\cite{cab_BLT1969}). We used this form in our previous works 
\cite{CR2005,CR2006,CRW2008,CRW2009},
where we considered only positive-energy particles. 

We can determine action of the operator defined in
Eq.~(\ref{spin_NW_def_C}) on basis vectors.
Taking into account Eqs.~(\ref{bispinor_rep_def},
\ref{unitary_rep_generators}, \ref{transformation_covariant},
\ref{Pauli-Lubanski_def}), and (\ref{pgamma_action}) we get in the
covariant basis
 \begin{equation}
 \hat{W}^\mu \ket{\alpha,\sg{\epsilon}p} = 
 - \frac{1}{2} \sg{\epsilon} \sum_\beta \big( ( \sg{\epsilon}m \gamma^\mu + p^\mu)\gamma^5
 \big)_{\alpha\beta} \ket{\beta,\sg{\epsilon}p}.
 \label{action_W_covariant}
 \end{equation} 
Furthermore, from Eqs.~(\ref{spin_NW_def_C}, \ref{action_W_covariant})
we receive 
 \begin{equation}
 \Hat{\vec{S}} \ket{\alpha,\sg{\epsilon}p} = 
 - \frac{1}{2} \Big[ \Big( \boldsymbol{\gamma} + 
 \frac{\sg{\epsilon}\vec{p}}{\sg{\epsilon}p^0+m}(I-\gamma^0)
 \Big) \gamma^5 \Big]_{\alpha\beta}
 \ket{\beta,\sg{\epsilon}p}.
 \end{equation}

Therefore, using
Eqs.~(\ref{transformation_covariant}, \ref{covariant_basis_inverse},
\ref{action_W_covariant}, \ref{formula_2}) we find in the spin basis
 \begin{equation}
 \hat{W}^\mu \ket{\sg{\epsilon}p,\sigma} = - \frac{m}{2} \sg{\epsilon} \sum_\lambda
 \big[ \bar{v}^{\sg{\epsilon}}(p) \gamma^\mu \gamma^5 v^{\sg{\epsilon}}(p)
 \big]_{\sigma\lambda} \ket{\sg{\epsilon}p,\lambda}.
 \label{action_W_spin}
 \end{equation}
Finally, by virtue of Eqs.~(\ref{formula_3}, \ref{formula_4}), we have
 \begin{gather}
 \hat{W}^0 \ket{\sg{\epsilon}p,\sigma} = \sum_\lambda \frac{1}{2}\sg{\epsilon}
 (\vec{p}\cdot\boldsymbol{\sigma}^T)_{\sigma\lambda} 
 \ket{\sg{\epsilon}p,\lambda}, \label{action_W0_spin}\\ 
 \Hat{\vec{W}} \ket{\sg{\epsilon}p,\sigma} = \sum_\lambda
 \frac{1}{2} \sg{\epsilon} \Big(  
 m \boldsymbol{\sigma}^T +
 \frac{\vec{p}(\vec{p}\cdot\boldsymbol{\sigma}^T)}{m+p^0} 
 \Big)_{\sigma\lambda} \ket{\sg{\epsilon}p,\lambda}.\label{action_Wi_spin}
 \end{gather}

Applying Eq.~(\ref{spin_NW_def_C}) we get
 \begin{equation}
 \hat{S}^i \ket{\sg{\epsilon}p,\sigma} = \frac{1}{2} \sum_\lambda
 (\boldsymbol{\sigma}^T)_{\sigma\lambda} \ket{\sg{\epsilon}p,\lambda},
 \label{spin_spin_basis}
 \end{equation}
which coincides with Eq.~(\ref{spin_action_spin_basis}).

As we have seen in Sec.~\ref{sec:FW} the Foldy-Woutheysen basis is in
fact the spin basis. In Ref.~\cite{FW1950} spin operator which after
Foldy-Woutheysen transformation has a form given in
Eq.~(\ref{spin_spin_basis}) was named 
``mean-spin operator''. Therefore, the spin operator defined in
Eq.~(\ref{spin_NW_def_abs}) 
coincides with Foldy-Woutheysen mean-spin.

We can define spin operator in yet another way, as a difference
between total angular momentum (which is defined with help of
Poincar\'e group generators $\hat{J}_{\alpha\beta}$ as $\hat{J}^i =
\frac{1}{2}\varepsilon_{ijk}\hat{J}^{jk}$) and orbital angular
momentum $\hat{\vec{X}}\times\hat{\vec{P}}$:
 \begin{equation}
 \hat{\vec{S}} = \hat{\vec{J}}-\hat{\vec{X}}\times\hat{\vec{P}}.
 \label{spin_orbital}
 \end{equation}
The Newton-Wigner position operator, discussed in
Sec.~\ref{sec:position}, can be expressed in terms of the
generators of the Poincar\'e group. In the case we consider here, 
i.e.\ when negative energies are allowed, the Newton-Wigner operator
takes the following form:
 \begin{equation}
 \hat{\vec{X}} = -\frac{1}{2}\Big(
 \frac{1}{\hat{P}^0} \hat{\vec{K}} + \hat{\vec{K}} \frac{1}{\hat{P}^0}
  \Big) - \frac{\hat{\vec{P}}\times\hat{\vec{W}}}{m \hat{P}^0
    (m+\hat{\sg{\mathcal{E}}}\hat{P}^0)},
 \label{NW_generators}
 \end{equation}
where $K^i=J^{0i}$ (compare \cite{TFJordan1980}).
Notice, that action of $\hat{\vec{J}}$ and $\hat{\vec{K}}$ on basis
vectors is independent of energy sign $\sg{\epsilon}$; action of
$\hat{P}^\mu$ and $\hat{\vec{W}}$ is given in
Eqs.~(\ref{four-momentum_eigenvectors}, \ref{action_W_covariant},
\ref{action_W0_spin}, \ref{action_Wi_spin}).
Inserting Eq.~(\ref{NW_generators}) into Eq.~(\ref{spin_orbital})
we can check that spin operator defined in this way also coincides
with (\ref{spin_NW_def_C}).

\subsection{Transformation properties of the spin operator}

In this section we find transformation properties of the spin operator
defined in Eq.~(\ref{spin_NW_def_positive}). To do this, let us
consider two inertial observers, $\mathcal{O}$ and
$\mathcal{O}^\prime$, connected by a Lorentz transformation
(\ref{Lorentz_transf}). The spin operator 
in the reference frame of the observer $\mathcal{O}$ is given by
Eq.~(\ref{spin_NW_def_positive}). We determine the form of this
operator in the frame of the observer $\mathcal{O}^\prime$ in terms of
the spin operator $\Hat{\vec{S}}$. 

Firstly, let
 \begin{equation}
 \Lambda(R) = \begin{pmatrix}
 1 & \vec{0}^T\\
 \vec{0} & R
 \end{pmatrix}, \qquad 
 R\in\textsf{SO}(3)
 \end{equation}
be a pure rotation. In this case we see immediately that
 \begin{equation}
 \Hat{\vec{S}}^\prime = R \Hat{\vec{S}},
 \label{spin_transf_rotation}
 \end{equation}
i.e.\ $\Hat{\vec{S}}$ transforms like a
vector under rotations---compare 
Eq.~(\ref{spin_transf_rotat_algebra}). Now, let 
$\Lambda(\vec{v})$ be a pure Lorentz boost.  In the frame
$\mathcal{O}^\prime$ the spin operator has obviously the following
form: 
 \begin{equation}
 \Hat{\vec{S}}^\prime = \frac{1}{m} \left( 
 \Hat{\vec{W}}^\prime - \Hat{W}^{\prime 0}
 \frac{\Hat{\vec{P}}^\prime}{\Hat{P}^{\prime 0}+m}
 \right),
 \label{spin_primed}
 \end{equation}
where 
 \begin{equation}
 \hat{W}^{\prime\mu} = \Lambda(\vec{v})^{\mu}_{\phantom{\mu}\nu} 
 \hat{W}^\nu, \qquad
 \hat{P}^{\prime\mu} = \Lambda(\vec{v})^{\mu}_{\phantom{\mu}\nu} 
 \hat{P}^\nu.
 \label{trans_for_spin}
 \end{equation}
The explicit form of the
most general pure boost is given in Eq.~(\ref{boost_general}).

Inserting Eqs.~(\ref{trans_for_spin}) into Eq.~(\ref{spin_primed}) and
using the following relations (yielded by
Eq.~(\ref{spin_NW_def_positive}) and $\Hat{P}^\mu \Hat{W}_\mu=0$):  
 \begin{align}
 \Hat{W}^0 & = \Hat{\vec{P}}\cdot\Hat{\vec{S}},\\
 \Hat{\vec{W}} & = m \Hat{\vec{S}} + (\Hat{\vec{P}}\cdot\Hat{\vec{S}})
 \frac{\Hat{\vec{P}}}{m+\Hat{P}^0},
 \end{align}
we get finally
 \begin{multline}
 \Hat{\vec{S}}^\prime = \Hat{\vec{S}} +
 \frac{(1-\gamma)(\Hat{\vec{P}}\cdot\Hat{\vec{S}})
 +\gamma(m+\Hat{P}^0)(\vec{v}\cdot\Hat{\vec{S}})}{(m+\Hat{P}^0)
 [m+\gamma(\Hat{P}^0-\vec{v}\cdot\Hat{\vec{P}})]} \Hat{\vec{P}}\\ 
 +\frac{\gamma}{m+\gamma(\Hat{P}^0-\vec{v}\cdot\Hat{\vec{P}})}\Big[  
 \frac{\gamma(m-\Hat{P}^0)(\vec{v}\cdot\Hat{\vec{S}})}{1+\gamma}\\
 + \frac{2\gamma(\vec{v}\cdot\Hat{\vec{P}})
 (\Hat{\vec{P}}\cdot\Hat{\vec{S}})}{(m+\Hat{P}^0)(1+\gamma)}
 -\Hat{\vec{P}}\cdot\Hat{\vec{S}}
 \Big]\vec{v},
 \label{spin_transf_final}
 \end{multline}
where $\gamma=(1-1/v^2)^{-1/2}$ is a Lorentz factor.
Now, comparing Eq.~(\ref{spin_transf_final}) with
Eq.~(\ref{Wigner_rot_final}) we see that 
 \begin{equation}
 \Hat{\vec{S}}^\prime = R(\vec{v},\Hat{P}) \Hat{\vec{S}},
 \label{spin_transf_Wigner}
 \end{equation}
where $R(\vec{v},p)$ is given in Eq.~(\ref{Wigner_rot_final}).
Thus, taking into account Eqs.~(\ref{spin_transf_rotation}) and
(\ref{spin_transf_final}), spin operator transforms under Lorentz
transformations according to Wigner rotation. 
Notice that $R(\vec{v},\Hat{P})$ in Eq.~(\ref{spin_transf_Wigner}) is
an operator. In momentum basis it is an ordinary matrix but in other
bases, like e.g.\ position basis, it is a non-local operator.

\section{Particle in electromagnetic field}

To make this paper self-contained we firstly remind here some results we
discussed in details in our paper \cite{CR2005}. Then we find the
transformation law for the Bloch vector describing fermion polarization. 

Let us consider a Dirac particle with positive energy
($\sg{\epsilon}=+1$) and sharp momentum $\vec{q}$. The most general
state of such a particle is described by the following density matrix:
 \begin{equation}
 \hat{\rho}(q,\boldsymbol{\xi}) =  
 \frac{1}{2}
 (1+\boldsymbol{\xi}\cdot\boldsymbol{\sigma})_{\sigma\lambda}
 \ket{q,\sigma}\bra{q,\lambda}, 
 \label{state_sharp}
 \end{equation}
where the Bloch vector $\boldsymbol{\xi}$ determines a polarization of
a particle. Using Eqs.~(\ref{action_W0_spin}, \ref{action_Wi_spin},
\ref{spin_spin_basis}) we can find the normalized average value of the
Pauli-Lubanski and spin operators in the state defined in
Eq.~(\ref{state_sharp}). We get
 \begin{align}
 \big< \hat{W}^0 \big>_{\hat{\rho}} & =
 \frac{\vec{q}\cdot\boldsymbol{\xi}}{2},\\ 
 \big< \hat{\vec{W}} \big>_{\hat{\rho}} & = \frac{1}{2} 
 \Big( m\boldsymbol{\xi} +
 \frac{\vec{q}(\vec{q}\cdot\boldsymbol{\xi})}{q^0+m}  \Big), \\ 
 \big< \hat{\vec{S}} \big>_{\hat{\rho}} & = \frac{\boldsymbol{\xi}}{2}.
 \end{align} 
Now, let us assume that a charged particle with sharp momentum moves
in the external electromagnetic field. The momentum and polarization
of such a particle can be regarded as functions of time
 \begin{equation}
 q=q(t),\qquad \boldsymbol{\xi}=\boldsymbol{\xi}(t).
 \end{equation}
The expectation value of the operators representing the spin and the
momentum will follow the same time dependence as one would
obtain from the classical Lorentz-covariant equations of motion
\cite{cab_Anderson1967,cab_BMT1959,cab_Corben1961,cab_CM1994}. The slow motion 
limit of the equations of motion, in the case when the electric field
is equal to zero, takes the form
 \begin{align}
 \frac{d\vec{q}}{dt} & = \frac{e}{m}\vec{q}\times\vec{B} +
 \frac{e}{2m} \boldsymbol{\xi}\cdot\nabla\vec{B},\\
 \frac{d\boldsymbol{\xi}}{dt} & = \frac{e}{m}
 \boldsymbol{\xi}\times\vec{B}, 
 \end{align}
(we assume that the giromagnetic ratio $g=2$). Therefore we should really
identify $\boldsymbol{\xi}$ with the polarization of a particle. 

Notice that the transformation law for the spin operator,
Eq.~(\ref{spin_transf_final}), is 
consistent with our identification. Indeed, the density matrix
$\hat{\rho}(q,\boldsymbol{\xi})$ given in Eq.~(\ref{state_sharp}) as
seen by the observer $\mathcal{O}^\prime$ has the form
 \begin{equation}
 \Hat{\rho}^\prime = U(\Lambda) \Hat{\rho}(q,\boldsymbol{\xi})
 U^\dagger(\Lambda) = \Hat{\rho}(\Lambda q,\boldsymbol{\xi}^\prime),
 \end{equation}
where
 \begin{equation}
 \boldsymbol{\xi}^\prime = R(\Lambda,q)\boldsymbol{\xi}. 
 \label{transf_xi}
 \end{equation}
To obtain the above relation we used 
the standard homomorphism of the
$\mathsf{SU}(2)$ group onto $\mathsf{SO}(3)$ group according to which 
 \begin{equation}
 \mathcal{D}(R(\Lambda,q))
 (\boldsymbol{\xi}\cdot\boldsymbol{\sigma})
 \mathcal{D}^\dagger(R(\Lambda,q))
 =
 \big(R(\Lambda,q)
 \boldsymbol{\xi}\big)\cdot\boldsymbol{\sigma}.
 \end{equation} 
Eq.~(\ref{transf_xi}) means that $\boldsymbol{\xi}$ transforms
according to the Wigner 
rotation. This is consistent with the transformation law for the spin
operator.

\section{Position representation}

Now, we want to define bispinors in a position representation and
vectors corresponding to them in an abstract Hilbert space. 
We are interested in a covariant picture therefore we use in our
construction the covariant basis [Eq.~(\ref{covariant_basis_def})].
Thus we define
 \begin{equation}
 \ket{x,\alpha,\sg{\epsilon}} = \frac{1}{(2\pi)^{3/2}} 
 \int \frac{d^3\vec{p}}{2\omega(\vec{p})} e^{-i\sg{\epsilon}px}
 \ket{\alpha,\sg{\epsilon}p},
 \label{vec_x_a_c}
 \end{equation}
and
 \begin{equation}
 \ket{x,\alpha} = \sum_{\sg{\epsilon}} \ket{x,\alpha,\sg{\epsilon}}. 
 \label{vec_x_a}
 \end{equation}
Inverse transformation reads
 \begin{align}
 \ket{\alpha,\sg{\epsilon}p} & = \frac{2m\sg{\epsilon}}{(2\pi)^{3/2}} \int d\sigma_\mu(x)
 e^{i\sg{\epsilon}px} \big(\Lambda_\sg{\epsilon}(p)\gamma^\mu\big)_{\alpha\beta} 
 \ket{x,\beta,\sg{\epsilon}}\nonumber\\
 & = \frac{2m\sg{\epsilon}}{(2\pi)^{3/2}} \int_{x^0=\text{const}} d^3\vec{x}
 e^{i\sg{\epsilon}px} \big(\Lambda_\sg{\epsilon}(p)\gamma^0\big)_{\alpha\beta} 
 \ket{x,\beta,\sg{\epsilon}},
 \end{align}
or, in terms of vectors (\ref{vec_x_a})
 \begin{align}
 \ket{\alpha,\sg{\epsilon}p} & = \frac{2m}{(2\pi)^{3/2}} \int d\sigma_\mu(x)
 e^{i\sg{\epsilon}px} \big(\Lambda_\sg{\epsilon}(p)\gamma^\mu\big)_{\alpha\beta} 
 \ket{x,\beta}\nonumber\\
 & = \frac{2m}{(2\pi)^{3/2}} \int_{x^0=\text{const}} d^3\vec{x}
 e^{i\sg{\epsilon}px} \big(\Lambda_\sg{\epsilon}(p)\gamma^0\big)_{\alpha\beta} 
 \ket{x,\beta}.
 \end{align} 

We can check, that
 \begin{equation}
 \bracket{\overline{\alpha,\sg{\epsilon}^\prime p}}{x,\beta,\sg{\epsilon}} = 
 \frac{\sg{\epsilon}\delta_{\sg{\epsilon}\sg{\epsilon}^\prime}}{(2\pi)^{3/2}} 
 e^{-i\sg{\epsilon}px} \big( \Lambda_{\sg{\epsilon}}(p) \big)_{\beta\alpha}. 
 \end{equation}

Now, bispinor in a position representation we define as follows:
 \begin{equation}
 \Psi_\alpha(x) = \bracket{\psi}{\alpha,x}.
 \end{equation}
Let us notice that, by virtue of
Eqs.~(\ref{bispinor_p_c},\ref{vec_x_a}), the above bispinor is 
related with bispinors in momentum representation via the standard
relation
 \begin{equation}
 \Psi_\alpha(x) = 
 \frac{1}{(2\pi)^{3/2}} \sum_{\sg{\epsilon}}
 \int \frac{d^3\vec{p}}{2\omega(\vec{p})}
 \psi^{\sg{\epsilon}}_{\alpha}(p) e^{-i\sg{\epsilon}px}.
 \end{equation}

One can also show that the scalar product (\ref{scalar_bispinor_p}) in
terms of $\Psi_\alpha(x)$ reads
 \begin{equation}
 (\psi,\phi)  = \int d^3\vec{x}\, \overline{\Psi}(x) \gamma^0 \Phi(x),
 \end{equation}
which is consistent with Eq.~(\ref{scalar_spinors_x0}).

It should be stressed here that vectors $\ket{x,\alpha,\sg{\epsilon}}$
defined in Eq.~(\ref{vec_x_a_c}) are not eigenvectors of the
Newton--Wigner position operator. 

It is possible to define Foldy--Wuotheysen transformations on the
level of bispinors in position representation. However, the connection
between bispinors before and after the Foldy--Woutheysen
transformation is non-local \cite{FW1950}.

\section{Conclusions}

In conclusion, we have formulated the Dirac formalism in an abstract
Hilbert space which is a carrier space of an unitary representation of
the Poincar\'e group. To include negative energy solutions of the
Dirac equation we have considered the direct sum of
carrier spaces of positive and negative energy unitary representations
of the Poincar\'e group for a massive spin-1/2 particle. 
We have introduced basis which under Lorentz transformations transforms
in a manifestly covariant manner according to the bispinor
representation of the Lorentz group. Vectors of the covariant basis in
a natural way fulfill the Dirac equation.
We have also shown that the Foldy-Woutheysen transformation which
diagonalizes the Dirac Hamiltonian, corresponds to the transformation
between covariant basis and the standard basis in the carrier space of
the Pioncar\'e group representation.

Moreover, we have discussed in detail the relativistic spin operator
for massive particle. We have shown that in the case of Dirac
particles the spin operator used in quantum field theory is
equal to the Foldy-Woutheysen mean-spin operator. We have also shown
that this spin operator under Lorentz group action transforms
according to the Wigner rotation matrix in which momentum is replaced
by momentum operator. Such a ``Wigner rotation'' is a highly non-local
operator.

\begin{acknowledgments}
This work has been supported by the University of Lodz and by the Polish
Ministry of Science and Higher Education under the contract
No.~N~N202~103738. 
\end{acknowledgments}

\appendix
\section{Basis in an abstract Hilbert space}
\label{App:basis} 

We discuss here in detail the choice of basis in an abstract Hilbert
space $\mathcal{H}=\mathcal{H}_+\oplus\mathcal{H}_-$.
Basis vectors should be labeled by four-momentum $p$, spin
$\sigma=\pm 1/2$, and index $\sg{\epsilon}=\pm 1$ identifying sign of energy.
Let us denote for a moment basis vectors as $\ket{p_{\sg{\epsilon}},\sigma}$.
By definition we have
 \begin{equation}
 p_{\sg{\epsilon}}^{0} = \sg{\epsilon} \omega(\vec{p}_{\sg{\epsilon}}),
 \end{equation}
where 
 \begin{equation}
 \omega(\vec{p}_{\sg{\epsilon}}) = + \sqrt{m^2+\vec{p}_{\sg{\epsilon}}}.
 \end{equation}
Furthermore, let $q_{\sg{\epsilon}}=(\sg{\epsilon}m,\vec{0})$ denote four-momentum
of a particle in its rest frame. 
We assume that vectors $\ket{p_{\sg{\epsilon}},\sigma}$ are generated from
$\ket{q_{\sg{\epsilon}},\sigma}$ in the following way:
 \begin{equation}
 \ket{p_{\sg{\epsilon}},\sigma} = U(L_{p_{\sg{\epsilon}}}^{\sg{\epsilon}}) 
 \ket{q_{\sg{\epsilon}},\sigma},
 \end{equation}
where $L_{p_{\sg{\epsilon}}}^{\sg{\epsilon}}$ is a standard Lorentz boost which fulfills
 \begin{equation}
 p_{\sg{\epsilon}} = L_{p_{\sg{\epsilon}}}^{\sg{\epsilon}} q_{\sg{\epsilon}},\qquad 
 L_{q_{\sg{\epsilon}}}^{\sg{\epsilon}} = I.
 \end{equation}
Standard Wigner induction method leads to the following form of the
Lorentz group action on vectors $\ket{p_{\sg{\epsilon}},\sigma}$:
 \begin{equation}
 U(\Lambda)\ket{p_{\sg{\epsilon}},\sigma} = 
 \mathcal{D}(R^{\sg{\epsilon}}(\Lambda,p_{\sg{\epsilon}}))_{\lambda\sigma}
 \ket{\Lambda p_{\sg{\epsilon}},\lambda}
 \end{equation}
where $\mathcal{D}$ denotes spin-1/2 representation of the rotation
group and
$R^{\sg{\epsilon}}(\Lambda,p_{\sg{\epsilon}})=(L_{\Lambda p_{\sg{\epsilon}}}^{\sg{\epsilon}})^{-1} 
 \Lambda L_{p_{\sg{\epsilon}}}^{\sg{\epsilon}}$ is a Wigner rotation.
We would like to have $\Lambda p_{\sg{\epsilon}} = (\Lambda p)_{\sg{\epsilon}}$. 
Therefore we take
 \begin{equation}
 p_{\sg{\epsilon}} = \sg{\epsilon} p = (\sg{\epsilon}\omega(\vec{p}),\sg{\epsilon}\vec{p}),
 \label{pc_choice}
 \end{equation}
and, consequently, basis vectors in the following form:
 \begin{equation}
 \ket{p_{\sg{\epsilon}},\sigma} = \ket{\sg{\epsilon}p,\sigma}.
 \end{equation}
Notice, that the choice which might seem to be the most natural, i.e. 
$p_{\sg{\epsilon}} = \sg{\epsilon}p^\pi=(\sg{\epsilon}\omega(\vec{p}),\vec{p})$ is not
suitable because $\Lambda p^\pi\not= (\Lambda p)^\pi$.

Moreover, we easily see that for the standard boost defined as
 \begin{equation}
 L_p = \begin{pmatrix}
 \tfrac{p^0}{m} & \tfrac{\vec{p}^T}{m} \\
 \tfrac{\vec{p}}{m} & I + \frac{\vec{p}\otimes\vec{p}^T}{m(m+p^0)}
 \end{pmatrix}
 \label{standard_boost}
 \end{equation}
we have
 \begin{equation}
 \sg{\epsilon} p = L_p q_{\sg{\epsilon}}.
 \end{equation}
Thus it holds
 \begin{equation}
 R^{\sg{\epsilon}}(\Lambda,\sg{\epsilon}p) = R(\Lambda,p) = L_{p}^{-1} \Lambda L_p,
 \end{equation}
and we finally receive
\begin{equation}
 U(\Lambda)\ket{\sg{\epsilon}p,\sigma} = 
 \mathcal{D}(R(\Lambda,p))_{\lambda\sigma}
 \ket{\sg{\epsilon}\Lambda p,\lambda}.
 \end{equation}

\section{Wigner rotation}

In this Appendix we give the explicit form of a Wigner rotation for a
Lorentz transformation $\Lambda$ being a pure boost. The most general
Lorentz boost $\Lambda(\vec{v})$ between two inertial frames of
reference, $\mathcal{O}$ and $\mathcal{O}^\prime$, 
 \begin{equation}
 x^{\prime \mu} = \Lambda(\vec{v})^{\mu}_{\phantom{\mu}\nu} x^\nu
 \end{equation}
can be written in the following form
 \begin{equation}
 \Lambda(\vec{v}) = \left(\begin{array}{c|c}
 \gamma & -\gamma \vec{v}^T \\
 \hline
 -\gamma\vec{v} & I+\frac{\gamma^2}{1+\gamma}\vec{v}\otimes\vec{v}^T 
 \end{array}\right), \label{boost_general}
 \end{equation}
where $\vec{v}$ is the velocity of a frame $\mathcal{O}^\prime$ with
respect to a frame $\mathcal{O}$ and $\gamma=(1-1/v^2)^{-1/2}$ is a
Lorentz factor.

Now, using Eqs.~(\ref{standard_boost}) and (\ref{boost_general}) we
can find by direct calculation 
 \begin{equation}
 R(\Lambda(\vec{v}),p) = L^{-1}_{\Lambda(\vec{v})p} \Lambda(\vec{v})
 L_p = \left(\begin{array}{c|c}
 1 & \vec{0}^T \\
 \hline
 \vec{0} & R(\vec{v},p)
 \end{array}\right),
 \end{equation}
where the matrix $R(\vec{v},p)\in\textsf{SO}(3)$ reads
 \begin{multline}
 R(\vec{v},p) = I + \frac{1-\gamma}{ab}\vec{p}\otimes\vec{p}^T + 
 \frac{\gamma^2(m-p^0)}{b(1+\gamma)}\vec{v}\otimes\vec{v}^T \\
 + \frac{\gamma}{b}\vec{p}\otimes\vec{v}^T + \frac{\gamma}{b}
 \Big( \frac{2\gamma(\vec{v}\cdot\vec{p})}{a(1+\gamma)}-1 \Big)
 \vec{v}\otimes\vec{p}^T,
 \label{Wigner_rot_final}
 \end{multline}
where
 \begin{align}
 a & = m+\Hat{P}^0, \\
 b & = m+\Hat{P}^{\prime 0} =
 m+\gamma(\Hat{P}^0-\vec{v}\cdot\Hat{\vec{P}}). 
 \end{align}

\section{Dirac matrices}\label{app:Dirac_matrices}

Dirac matrices fulfill the relation
$\gamma^\mu\gamma^\nu+\gamma^\nu\gamma^\mu=2g^{\mu\nu}$ where
the Minkowski metric tensor $g^{\mu\nu}=\text{diag}(1,-1,-1,-1)$;
moreover we adopt the convention $\varepsilon^{0123}=1$.  We use the
following explicit representation of gamma matrices:
 \begin{equation}
 \gamma^0=\left(\protect\begin{array}{cc}
 0 & I \\ I & 0
 \end{array}\right),\quad 
 {\boldsymbol{\gamma}}=\left(\protect\begin{array}{cc}
 0 & -{\boldsymbol{\sigma}} \\ {\boldsymbol{\sigma}} & 0
 \protect\end{array}\right),\quad \gamma^5= 
 \left(\protect\begin{array}{cc} 
 I & 0 \\ 0 & -I
 \end{array}\right),
 \label{gamma_explicit}
 \end{equation} 
where $\boldsymbol{\sigma}=(\sigma_1,\sigma_2,\sigma_3)$ and
$\sigma_i$ are standard Pauli matrices.

\section{Useful formulas}
\label{sec:formulas}

The explicit form of amplitudes $v^\sg{\epsilon}(p)$ is the following:
\begin{equation}
 v^{\sg{\epsilon}}(p) = \frac{1}{2\sqrt{1+\frac{p^0}{m}}} 
 \begin{pmatrix}
 I_2 + \frac{1}{m} p^\mu \sigma_\mu \\[1mm]
 \sg{\epsilon} (I_2 + \frac{1}{m} {p^\pi}^\mu \sigma_\mu)
 \end{pmatrix} \sigma_2,
 \label{amplitudes_explicit}
 \end{equation}
where $\sigma_0=I_2$. It holds
 \begin{gather}
 \bar{v}^{\sg{\epsilon}}(p) \gamma^\mu v^{\sg{\epsilon}}(p) = \frac{p^\mu}{m} I_2,
 \label{formula_1}
 \\
 \bar{v}^{\sg{\epsilon}}(p) \gamma^5 v^{\sg{\epsilon}}(p) = 0.
 \label{formula_2}
 \end{gather}
Using Eqs.~(\ref{gamma_explicit}, \ref{amplitudes_explicit}) we find
 \begin{gather}
 \bar{v}^{\sg{\epsilon}}(p) \gamma^0 (\vec{p}\cdot\boldsymbol{\gamma})
 v^{\sg{\epsilon}}(p) = 0. 
 \label{formula_5}\\
 \bar{v}^{\sg{\epsilon}}(p) \gamma^0 \gamma^5 v^{\sg{\epsilon}}(p) = - \frac{1}{m}
 (\vec{p}\cdot\boldsymbol{\sigma}^T),
 \label{formula_3}\\
 \bar{v}^{\sg{\epsilon}}(p) \boldsymbol{\gamma} \gamma^5 v^{\sg{\epsilon}}(p) = -
 \frac{1}{m} \Big(  
 m \boldsymbol{\sigma}^T +
 \frac{\vec{p}(\vec{p}\cdot\boldsymbol{\sigma}^T)}{m+p^0} 
 \Big).
 \label{formula_4}
 \end{gather}

%

\end{document}